\documentclass[twocolumn,10pt,a4paper]{article}

\usepackage{graphicx,subfigure}
\usepackage{amsmath,amssymb,amsfonts}
\usepackage{units}
\usepackage{placeins}
\usepackage[normalleading]{savetrees}

\begin{document}

\title{The RMS Charge Radius of the Proton and Zemach Moments}

\author{Michael O.~Distler\footnote{e-mail distler@kph.uni-mainz.de}, 
Jan C.~Bernauer\footnote{now at: Laboratory for Nuclear Science, MIT, 
Cambridge,  Massachusetts 02139, USA} and Thomas Walcher\\
\large Institut f\"{u}r Kernphysik\\ Johannes-Gutenberg-Universit\"{a}t Mainz}
\maketitle

\begin{abstract}
On the basis of recent precise measurements of the electric form
factor of the proton, the Zemach moments, needed as input parameters
for the determination of the proton rms radius from the measurement of
the Lamb shift in muonic hydrogen, are calculated. It turns out that
the new moments give an uncertainty as large as the presently stated 
error of the recent Lamb shift measurement of Pohl {\it et al.}.
De~R\'{u}jula's idea of a large Zemach moment in order to reconcile
the five standard deviation discrepancy between the muonic Lamb shift
determination and the result of electronic experiments is shown to be
in clear contradiction with experiment. Alternative explanations are
touched upon.
\end{abstract}


\section{Introduction} \label{Intro}
Recently, two new precise measurements of the proton root mean square
(rms) radius have been published which deviate by 5 standard
deviations. The first one studying muonic hydrogen resulted after a
ten years effort in
$r_p = \sqrt{\langle r^2 \rangle} = \unit[0.84184(67)]{fm}$
\cite{Pohl:2010zz}, a factor of about 10 more precise
than all previous determinations. These previous determinations are
based on three very different methods: The CODATA values are derived
mostly from electronic hydrogen giving $r_p=\unit[0.8768(69)]{fm}$
\cite{Mohr:2008fa}, the Lamb shift of electronic hydrogen resulting in
$r_p=\unit[0.883(14)]{fm}$ \cite{Udem:1997zz,Melnikov:1999xp}, and
electron scattering from hydrogen yielding $r_p=\unit[0.895(18)]{fm}$
\cite{Sick:2003gm}. These ``electronic'' determinations were recently
corroborated by a new precise determination from electron scattering
at the Mainz Microtron MAMI giving $r_p=\unit[0.879(8)]{fm}$
\cite{Bernauer:2010wm}. Even considering the larger error of the
electronic determinations the deviation of $\approx\unit[0.04]{fm}$
has a significance of five standard deviations. This unexpected result
created quite a stir in the hadron community and asks for explanation.

The following proposals have been ventilated \cite{DeRujula:2010dp}:
\begin{itemize}
\item {\it The experimental results are not right.}

This is highly unlikely. The electronic experiments are based on well
established methods and the agreement of their results cannot be
accidental. The mu\-on\-ic experiment represents a very sophisticated
study of a group beyond any doubts.

However, in a followup paper to ref.~\cite{DeRujula:2010dp}
De~R\'{u}jula \cite{DeRujula:2010zk} questions just the correctness of
the electron-\-proton scattering (ep) results and a refutation of his
conjectures is one of the aims of this paper.

\item {\it The relevant QED calculations are incorrect.}

It is true that all determinations with electro\-mag\-netic probes
require higher order corrections, mostly called ``radiative
corrections''. However, in the case of the electronic determinations
the methods used are completely different requiring as different
corrections. This is particularly true for the about half dozen
electron scattering experiments performed in different laboratories
at very different kinematic conditions. The agreement of the
electronic experiments cannot be by chance.

The QED calculations, on which the Lamb shift determination in muonic
hydrogen is based, are very detailed and sophisticated and have been
improved again and again over more than four decades. However, the
perturbative expansion requires the knowledge of higher order radial
moments of the charge distribution, so called Zemach moments
\cite{Zemach:1956zz,Friar:1978wv}. In a recent paper De~R\'{u}jula
argued that the solution of the problem might be found in a third
Zemach moment as large as $\langle r^3 \rangle_{(2)} =
\unit[36.6]{fm^3}$ \cite{DeRujula:2010dp}. Immediately after that
publication Clo\"{e}t and Miller \cite{Cloet:2010qa}
showed that the proposal of De~R\'{u}jula is in contradiction with
experiment. Nevertheless, De~R\'{u}jula insisted on his arguments by
suggesting that the experimental basis is wrong
\cite{DeRujula:2010ub,DeRujula:2010zk}. We shall show on the basis of
the most recent, precise experimental results of Bernauer {\it et al.}\ 
\cite{Bernauer:2010wm} that this is not true and this is the central
subject of this paper.

It is noteworthy that QED considerations show
\cite{Bawin:2000px,Jentschura:2010} that the electron scattering
experiments and the atomic hydrogen spectroscopy determine the same
radius.

\item {\it There is, at extremely low energies and at the level of
accuracy of the lepton-proton atom experiments, ``physics beyond the
standard model''.}

After having excluded the previous point there stays the need for an
explanation which we will briefly discuss in the conclusions. However,
it is the conviction of the present authors that one has to
investigate more ideas within the standard model before one may
venture in physics beyond it.

\item {\it A single-dipole form factor is not adequate to the analysis
of precise low-energy data.}

This is definitely true as we shall show in this paper. However, the
small deviations from the standard one-dipole form \cite{Bernauer:2010wm}
will change the input into the determination of the muonic Lamb shift
only very mildly, but are at dramatic variance with the proposal of
De~R\'{u}jula \cite{DeRujula:2010dp}.
\end{itemize}

\section{Summary of our knowledge
  of the electric form factor of the proton}\label{sec:2}

Before we can turn to the central point, we have to summarise
our knowledge about the electric form factor of the proton
$G_{E,p}$. This knowledge started with the key paper of Janssens
{\it et al.}~\cite{Janssens:1965kd} using the Rosenbluth separation
of the electric and magnetic form factors in electron scattering.
They found as the shape of $G_{E,p}$ what is called today the standard
dipole form $G_{E,p} = 1/[1+Q^2/(\unit[0.71]{(GeV/c)^2})]^2$. After quite
a few further determinations this shape stayed the accepted reference
until about the year 2000. It was a great surprise that this reference
had to be modified after experiments performed at Jlab using polarised
electrons with polarised targets or by measuring the polarisation transfer
to the recoil proton. They showed a dramatic faster decline than the
standard dipole form at large $Q^2$. Several reviews have been published
on this topic in the last decade
\cite{HydeWright:2004gh,Perdrisat:2006hj,Arrington:2007ux}
giving access to the extended literature.

At low momentum transfer squared $Q^2$ a particularly precise
determination could be reached recently by fitting a large variation
of form factor models to the measured differential cross sections
directly, i.e.\ extending the Rosenbluth method
\cite{Bernauer:2010wm}.  This determination showed some structure at
scales of the pion cloud and it is this scale which is considered to
be particularly essential for the correct determination of the Zemach
moments. However, as we will show, the form factor shape is important
at all scales due to the features of the Fourier transform. It is
needed for the derivation of the charge distribution, indispensable
for the calculation of the Zemach moments from the measured form
factors. It turns out that the low $Q^2$ shape influences the
precision of the Zemach moments, but the bulk behaviour is essentially
insensitive to the specific model chosen.

For large $Q^2$ we use the results based on the fit of the pre-2007
data by Arrington {\it et al.}~\cite{Arrington:2007ux}, however, without
two photon corrections. The possible influence of these still unsafe
corrections is small in the relevant $Q^2$ range and does not change the
results in this paper.

\begin{figure}[ht]
\begin{center}
\includegraphics[width=0.9\columnwidth]{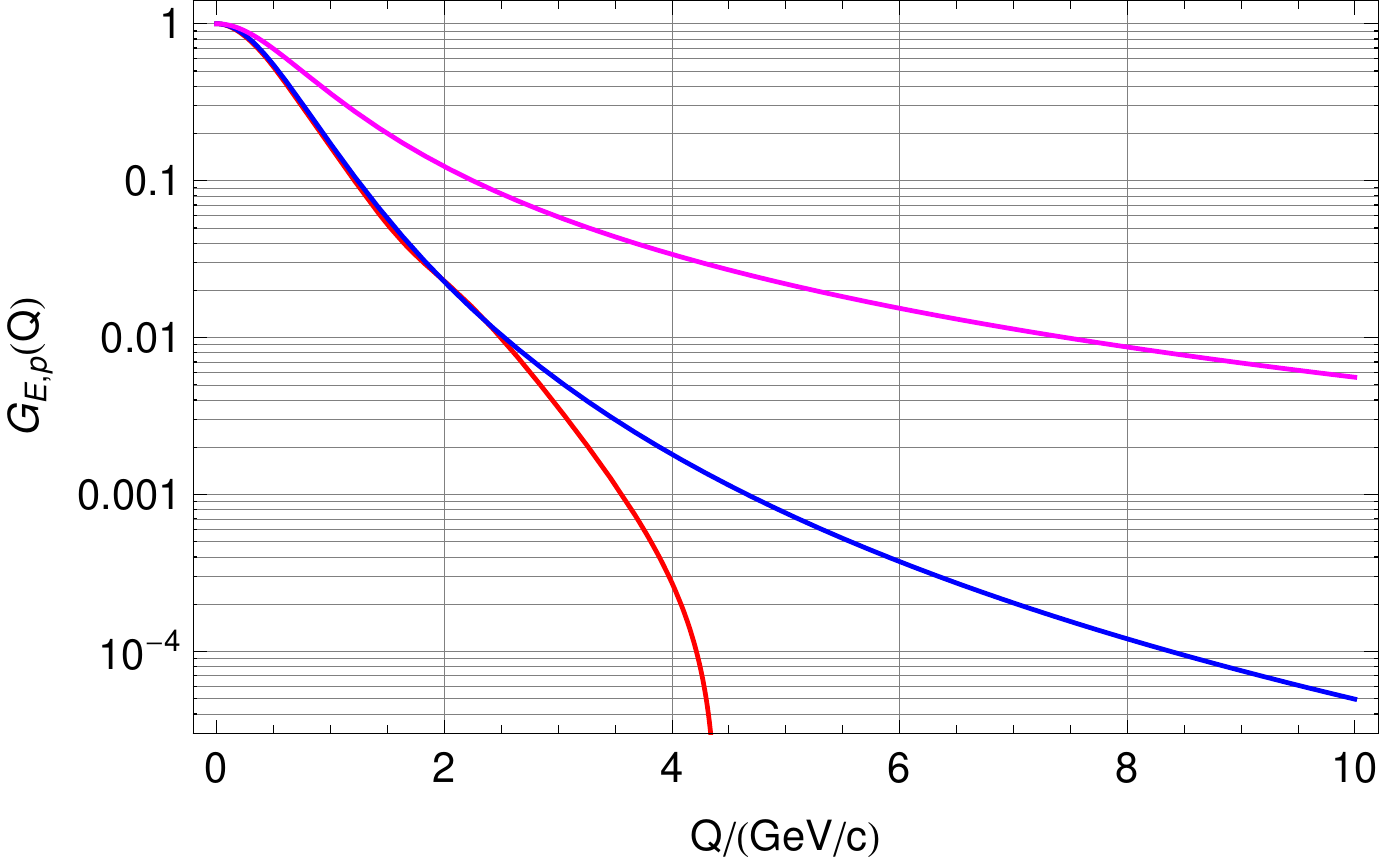}
\end{center}
\caption{The electric form factor of the proton $G_{E,p}$ as given by
the standard dipole form (blue curve) and as given by the assembly of the
fits of Bernauer {\it et al.}\ and Arrington {\it et al.}\ (red curve). The
toy model of De~R\'{u}jula (magenta curve) will be dealt with in
section~\ref{sec:4}. For details see text.}
\label{fig:1}
\end{figure}

Figure~\ref{fig:1} shows the standard dipole form and the shape used
in this paper. This shape consists of the Bernauer {\it et al.}\ results
for $Q<\unit[0.7]{GeV/c}$ and of the fits of Arrington {\it et
al.}\ for $Q\geqslant\unit[0.7]{GeV/c}$. From the many models of
Bernauer {\it et al.}\ which are equivalent within the statistics, we
have taken the standard dipole times a power series of order 8 to
produce the plots and the numerical results. If errors are quoted
for the Zemach moments given in section \ref{sec:3} the full
spread of models is considered with and without readjustment of the
Arrington {\it et al.}\ normalisations by as much as 2\%. In the future,
a new global fit including all data is highly desirable, however, as checks
show, for the purpose of  this letter no better information is needed.

\begin{figure}[ht]
\begin{center}
\includegraphics[width=0.9\columnwidth]{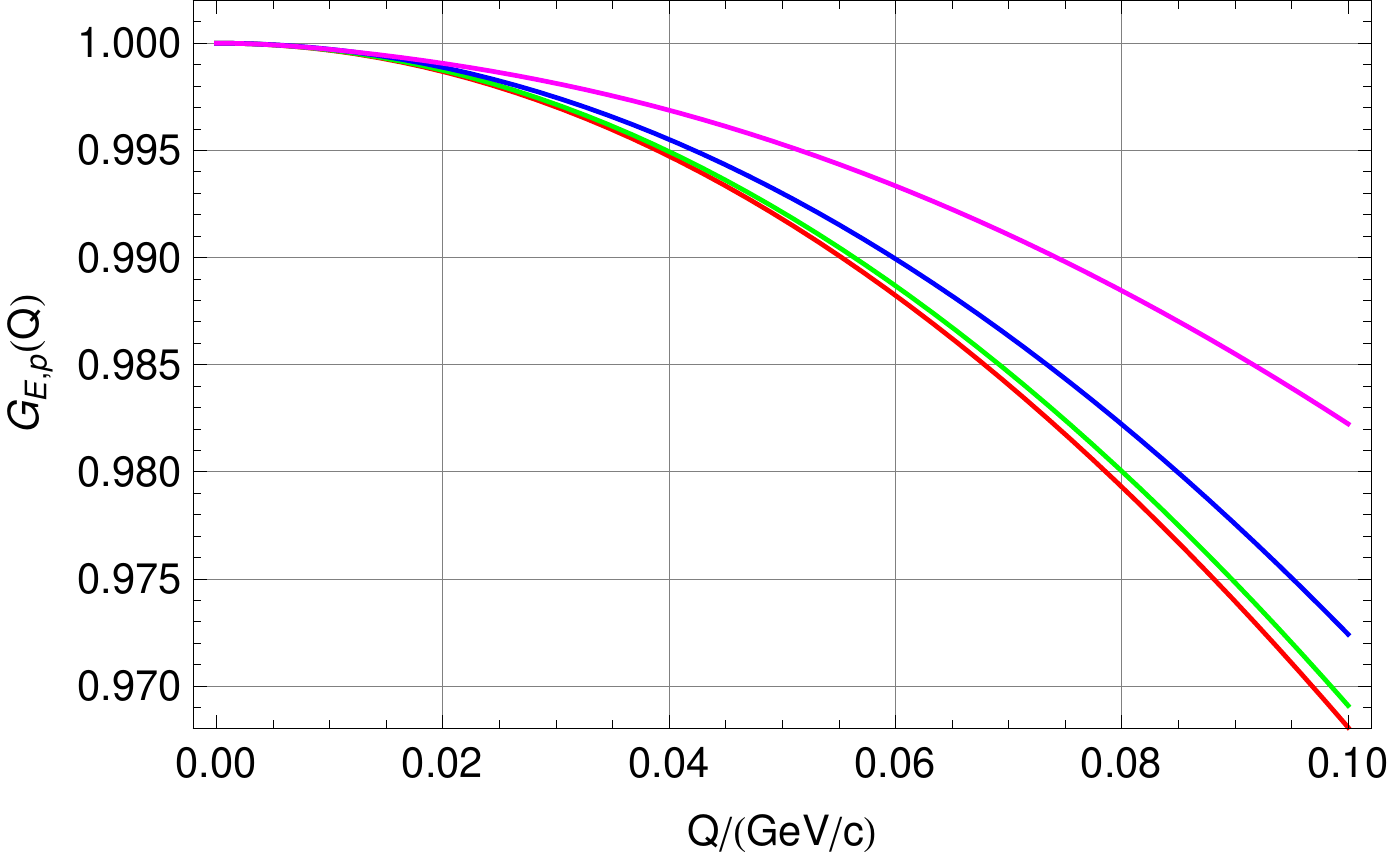}
\end{center}
\caption{As Fig.~\ref{fig:1} but in linear scale for small $Q$. The
red curve depicts the Bernauer {\it et al.}\ form and the green that of
Arrington {\it et al.}. The blue curve is again the standard dipole and
the magenta curve the toy model of De~R\'{u}jula.}
\label{fig:2}
\end{figure}

The shapes of Bernauer {\it et al.}\ and Arrington {\it et al.}\ differ
a few per mill at small $Q$ as shown in the linear plot in
Fig.~\ref{fig:2}.

With these form factor parametrisations we can now determine the
Zemach moments needed for the calculation of the hyperfine splitting
and the Lamb shift in hydrogen \cite{Zemach:1956zz,Friar:1978wv}.

\FloatBarrier

\section{Zemach moments} \label{sec:3}

The Zemach moments of the nuclear charge distribution come about
through the smearing of the Coulomb potential with the extended charge
distribution and the perturbative expansion of the hydrogen wave
functions \cite{Zemach:1956zz,Friar:1978wv}. They are defined by

\begin{equation}
\langle r^n \rangle_{(2)} = \int d^3r \,r^n \rho_{(2)}(r)
\label{eq:1}
\end{equation}

where $\rho_{(2)}(r)$ is the convolution of the charge or magnetic
distribution for the Lamb shift or hyperfine interaction,
respectively:
\begin{multline}
\rho_{(2)}(r)= \\
\int d^3r_2 \, \rho_{\textrm{charge}}(|\vec{r}-\vec{r_2}|) \,
\rho_{\textrm{charge or magnetic}}(r_2).
\label{eq:2}
\end{multline}

Since we are at this point concerned with the Lamb shift, the proton
charge distribution is folded with itself for the calculation of
$\rho_{(2)}(r)$. Inserting the Fourier transform of the Sachs electric
form factor $G_E(Q^2)$ for $\rho$ in Eq.~(\ref{eq:2}), and integrating
repeatedly by parts, one finds that the first and the third Zemach
moment can also be expressed in momentum space
\cite{Pachucki:1996zza}:

\begin{eqnarray}
\langle r^1 \rangle_{(2)} \!\!\! &=& \!\!\!
-\frac{4}{\pi} \, \int_0^{\infty} \frac{dQ}{Q^2}\left(G_E^2(Q^2)-1\right)
 \label{eq:3} \\[2ex]
\langle r^3 \rangle_{(2)} \!\!\! &=& \!\!\!
\frac{48}{\pi} \, \int_0^{\infty} \frac{dQ}{Q^4}
\left(G_E^2(Q^2)-1+\frac{Q^2}{3}\langle r^2 \rangle\right).
\label{eq:4}
\end{eqnarray}

These integrals can be evaluated analytically for the dipole form and
the monopole form (see section~\ref{sec:4}). For the general form
factors of section~\ref{sec:2} we integrate Eqs.~(\ref{eq:3}) and
(\ref{eq:4}) numerically.

\begin{figure}[ht]
\begin{center}
\includegraphics[width=0.9\columnwidth]{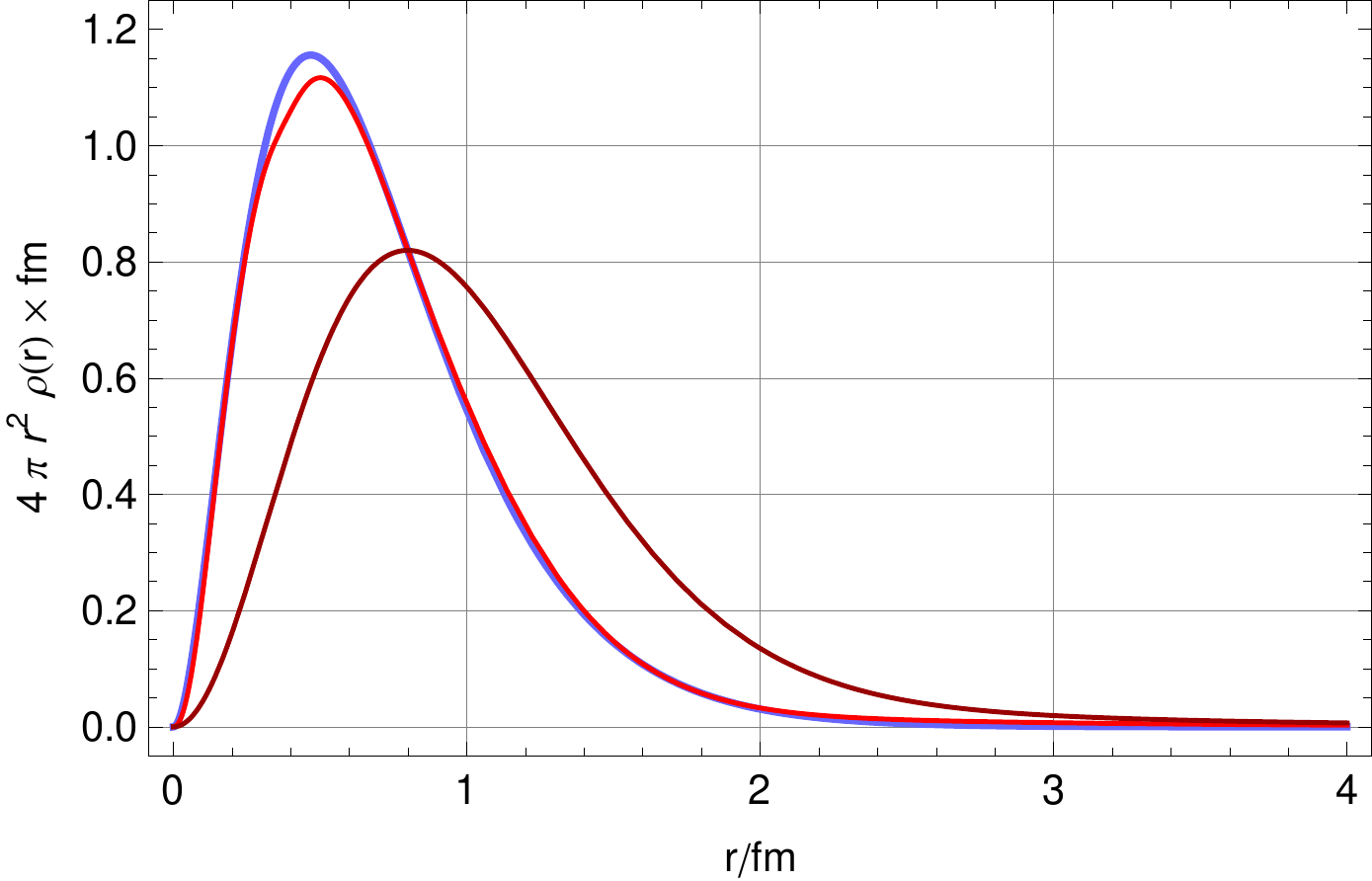}
\end{center}
\caption{The charge distribution $\rho_{\textrm{charge}}(r)$ (red
curve), the folded distribution (dark red curve) $\rho_{(2)}(r)$ for
the assembly of the $G_{E,p}$ fits of Bernauer {\it et al.}\ and
Arrington {\it et al.}\ (red curve in Fig.~\ref{fig:1}), and the
standard dipole distribution (blue curve) are shown. The stability of
the numerical method used to perform the Fourier transform can be seen
from the flatness of the curve at larger radii. The shift of charge
from low radii to the tail (of the light red curve) can be interpreted
as ``pion cloud'', however, such an interpretation needs a careful
qualification \cite{Vanderhaeghen:2010nd}.}
\label{fig:3}
\end{figure}

For illustration, we show in Fig.~\ref{fig:3} the charge
distribution $\rho_{\textrm{charge}}(r)$ and the folded distribution
$\rho_{(2)}(r)$ for the assembly of the $G_{E,p}$ fits of Bernauer
{\it et al.}\ and Arrington {\it et al.}\ depicted in Fig.~\ref{fig:1}.

It is instructive to study the integrand of Eq.~(\ref{eq:4}) and
learn where $G_E^2(Q^2)$ contributes. The third Zemach moment
calculated from the assembly of Bernauer {\it et al.}\ and Arrington
{\it et al.}\ fits is a factor 1.4 larger than the standard dipole
reference. For this form about 60\% of the integral arise from
momentum transfers $Q<\unit[0.7]{GeV/c}$, but 75\% of the surplus
originate in this interval and 50\% from $Q<\unit[0.3]{GeV/c}$.
This demonstrates the particular importance of low $Q$ electron
scattering data for the evaluation of the Zemach moments, an
observation already rightly emphasized by De~R\'{u}jula in
Ref.~\cite{DeRujula:2010ub}.

\begin{table*}[ht]
\caption{Moments of the charge distribution and Zemach moments of the
respective parametrizations of the form factor $G_{E,p}$. All
distribution functions are duly normalising to one. The ratio $f$ has
been used in Ref.~\cite{Pohl:2010zz} for the approximation of the
third Zemach moment. Errors for the Bernauer-Arrington fit assembly
have been determined using the full spread of the form factor fits in
Ref.~\cite{Bernauer:2010wm}. The moments for De~R\'{u}jula's toy model
were calculated with parameters typical for the CODATA rms radius:
$M=\unit[0.750]{GeV/c^2}$, $m=\unit[0.020]{GeV/c^2}$, and
$sin^2(\theta) = 0.3$.}
\label{tab:1}
\begin{center}
\begin{tabular}{l|r@{.}lr@{.}lr@{.}lr@{.}l|r@{.}l}
$G_{E,p}$ &  \multicolumn{2}{c}{$\langle r^2 \rangle \,/\,\hbox{fm}^2$}
         &  \multicolumn{2}{c}{$\langle r^1 \rangle_{(2)} \,/\,\hbox{fm}$}
         &  \multicolumn{2}{c}{$\langle r^2 \rangle_{(2)} \,/\,\hbox{fm}^2$}
         &  \multicolumn{2}{c|}{$\langle r^3 \rangle_{(2)} \,/\,\hbox{fm}^3$}
         &  \multicolumn{2}{c}{$f=\langle r^3 \rangle_{(2)}/\langle r^2 \rangle^{3/2}$}\\[1ex]\hline
Standard dipole \cite{Janssens:1965kd}
         & 0&6581    & 1&0246    & 1&316    & 2&023         & 3&789    \\
Friar-Sick \cite{Friar:2005jz}
         & 0&801(36) & \multicolumn{4}{c}{} & 2&71(13)      & 3&78(31) \\
Arrington \cite{Arrington:2007ux}
         & 0&742     & 1&077     & 1&484    & 2&50          & 3&91     \\
Bernauer-Arrington \cite{Bernauer:2010wm,Arrington:2007ux}
         & 0&774(8)  & 1&085(3)  & 1&553(16)& 2&85(8)       & 4&18(13) \\[0.5ex]
\hline
Toy model& 0&771     & 0&807     & 1&542    & 36&2          & 53&5     \\
De~R\'{u}jula \cite{DeRujula:2010dp}
         & 0&7687    & \multicolumn{4}{c}{} & 36&6(7.7)     & 52&2
\end{tabular}
\end{center}
\end{table*}

By numerical integration of the form factors according to
Eqs.~(\ref{eq:3}) and (\ref{eq:4}) and the folded distribution in
Eq.~(\ref{eq:1}) we get the moments listed in Table~\ref{tab:1}. A
good cross check of the numerical accuracy is the fulfillment of the
identity $\langle r^2 \rangle_{(2)}=2\langle r^2 \rangle$
\cite{Zemach:1956zz}.  For the moments in Table~\ref{tab:2} the
weighted charge distribution derived by numerical Fourier transform of
the Arrington-\-Bernauer parametrisation is integrated
numerically. Where possible the calculations were compared to
evaluations in momentum space, e.g.\ $\langle r^2 \rangle$ can also be
determined from the slope of $G_E$ at $Q^2=0$. While the agreement is
very good for the moments of lower order, we find that for higher
orders the dependence on the form factor models gets more important
and the numerical stability gets worse resulting in larger errors.
For $\langle r^5 \rangle$ we estimate a numerical uncertainty
comparable to the statistical error and for $\langle r^6 \rangle$ this
uncertainty is as big as 1.7 times the statistical error.

The rms radius is derived from the measured Lamb shift via the numerical
equation \cite{Borie:2004fv}
\begin{multline}
L^{5th}[\langle r^2 \rangle,\langle r^3 \rangle_{(2)}] = \\
\left( 209.9779(49)- 5.2262\,\frac{\langle r^2 \rangle}{\hbox{fm}^2} +
0.00913\,\frac{\langle r^3 \rangle_{(2)}}{\hbox{fm}^3}\right)\textrm{meV}.
\label{eq:5}
\end{multline}
showing that in particular the moments $\langle r^2 \rangle$ and
$\langle r^3 \rangle_{(2)}$ are of interest here.
For the Zemach moment $\langle r^3 \rangle_{(2)}$, unknown without the
knowledge of the charge distribution, the approximation $\langle r^3
\rangle_{(2)} \approx f \langle r^2 \rangle^{3/2}$ has been
adopted in Ref.~\cite{Pohl:2010zz}. The factor $f$ used in the
extraction of the muonic rms radius was $f=3.79$, i.e.\ in accord
with that of the standard dipole and that of Friar and Sick (see
Table~\ref{tab:1}). The new precise experiment of Bernauer
{\it et al.}\ yields an improved value of $f=4.18(13)$. This new
value would increase the $r_p$ in Ref.~\cite{Pohl:2010zz}
by +0.00025\,fm -- only a minor correction within the margin of the
quoted theoretical error. One could be tempted to use the new
experimental value for the third Zemach moment $\langle r^3
\rangle_{(2)}$ and ``improve'' the muonic result to give
$r_p=\unit[0.84245(67)]{fm}$, but this would be somewhat inconsistent
as we insist on the larger radius and the corresponding form factor
obtained from electron scattering. If one would interpret this
difference as a lack of knowledge of $\langle r^3 \rangle_{(2)}$ one
had to add it as a systematic error in linear and double the error in
Ref.~\cite{Pohl:2010zz}.

Although not essential for the purpose of this paper, but of general
interest in atomic physics, we have calculated additional moments of
the proton electro\-magnetic distributions. The moments listed in
Table~\ref{tab:2} are required for calculating the higher order finite
size corrections to the Lamb shift and to the hyperfine splitting in
electronic and muonic hydrogen. Some of the values differ
significantly from the standard dipole approximation.


\section{De~R\'{u}jula's toy model} \label{sec:4}

We now turn to the ``toy'' model of De~R\'{u}jula
\cite{DeRujula:2010dp} designed to explain the difference between the
electronic results and the muonic determination.

\begin{figure}[hbt]
\begin{center}
\includegraphics[width=0.9\columnwidth]{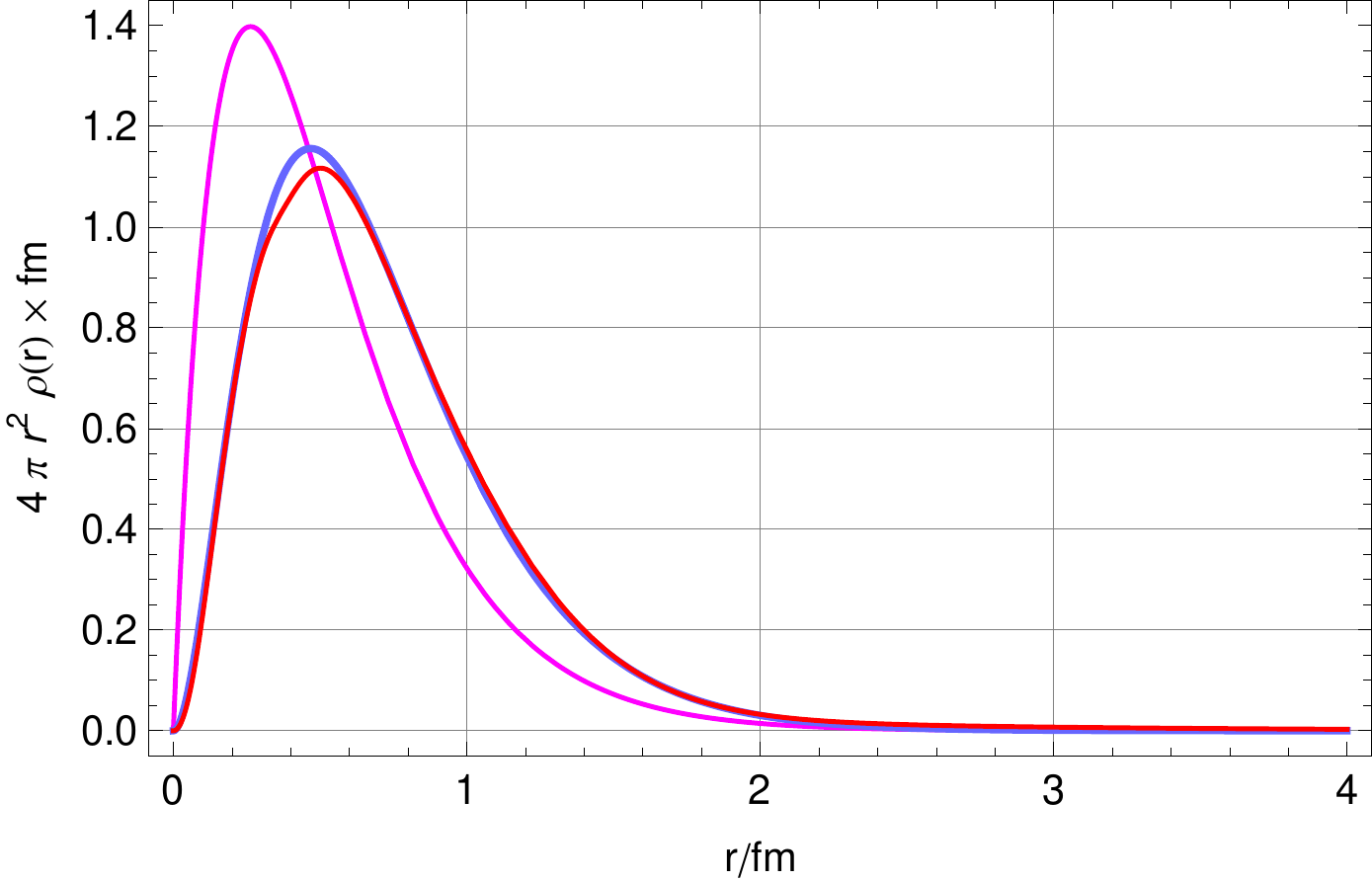}
\end{center}
\caption{The charge distribution derived from the Fourier transform of
the Bernauer-Arrington parametrisation of $G_{E,p}$ (red curve) and
the charge distribution of the toy model of De~R\'{u}jula
\cite{DeRujula:2010dp} (magenta curve). The De~R\'{u}jula charge
distribution was calculated with parameters given in the caption of
Table~\ref{tab:1}. For comparison the charge distribution of the
standard dipole form factor is shown (blue curve).}
\label{fig:4}
\end{figure}

The basic, very physical idea is the assumption of a charge cloud
reaching out to very large scales of $\approx\unit[20]{fm}$.
De~R\'{u}jula designs the charge distribution in his toy model in such
a way that the rms radius is constrained by the rms radii of Sick
\cite{Sick:2003gm} or that of CODATA \cite{Mohr:2008fa}. The third Zemach
moment is made so large, i.e.\ $\langle r^3 \rangle = \unit[36.6]{fm^3}$,
that the rms radius derived from the muonic Lamb shift (see Eq.~(\ref{eq:5}))
would be in accord with the electronic results. This increase is accomplished
by adding to a ``Yukawian charge distribution'', corresponding to a monopole
form factor, with short range, an exponential distribution, corresponding to
a dipole form factor, with long range. As illustration Fig.~\ref{fig:4} shows
one typical De~R\'{u}jula charge distribution compared to the distribution
derived by Fourier transform from the measured $G_{E,p}$ presented in
section~\ref{sec:2}.

The two charge distributions are in dramatic disagreement. In order to
make the disaccord more visible for the experimentally observable
$G_{E,p}$ we show in Fig.~\ref{fig:1} the Fourier transform of the
De~R\'{u}jula charge distribution compared to the $G_{E,p}$ of the
Bernauer-Arrington pa\-ram\-e\-tri\-sa\-tion and the standard dipole
form.  It is evident that the De~R\'{u}jula model is
excluded by experiment. The same was already shown on the basis of
older data and fits by Clo\"{e}t and Miller \cite{Cloet:2010qa}.

The basic problem of the De~R\'{u}jula argumentation is the design of
his model by the constraint of two range momenta only, instead of
using the full information available in the form factors. The scale
arguments are easily leading astray since scales define wave
trains. One needs the full momentum spectrum in a Fourier transform to
describe an object of finite extent. In his most recent followup
paper~\cite{DeRujula:2010zk} De~R\'{u}jula concluded with the request
{\it \ldots it would be very helpful to extract the correlation
dictated by the ep data \ldots} and this is exactly provided in this
paper, however, in the complete way of form factors going much beyond
the correlations of just two moments.

Not being able to maintain his large third Zemach moment on the basis of
experiment he takes the attitude that the experiments are wrong or at
least not sufficiently precise to exclude his conjectures
\cite{DeRujula:2010zk}. Of the several inaccuracies in his arguments
we have to rectify three:
\begin{itemize}
\item The {\it dark trumpet} in Fig.~1 of Bernauer {\it et
al.}~\cite{Bernauer:2010wm} includes all errors and represents
new, precise ep cross section data. The extrapolation to $Q^2 =0$
determining $\langle r^2 \rangle$ is much less arbitrary than
insinuated in Fig.~2 of Ref.~\cite{DeRujula:2010zk} since the red
arrows neglect the normalisation $G_E(Q^2=0)=1$ imposed by the total
charge of the proton. Since no experiment could determine the
cross sections with an absolute precision better than a few
percent, they are normalised to $G_E(Q^2=0)=1$ using a certain
analytical hypothesis for the form factor shape. As for the
determination of the rms radius from the small $Q^2$ behaviour this
has been amply discussed in Ref.~\cite{Sick:2003gm}.
\item The p-value of the $\chi^2$ distribution is utilised as an
important argument at several places in the paper, however, its
meaning is not correctly applied. In short: A $\chi^2 \gg
n_{dof}$ means either that the fit model is insufficient and/or that
the data sample has larger than statistical errors. Almost all
experiments have systematic errors and by increasing the errors in
Ref.\ \cite{Bernauer:2010wm} by as little as 7\% one would get a
$\chi^2/n_{dof} =1$ indicating a small systematic error indeed. The
issue concerns the difference between the frequentist and the Bayesian
interpretation of a fit. Additionally, in hypothesis testing the
p-value does not give the probability that the null hypothesis is
true. A good reference is the CERN classic of Frederick James
\cite{James:2006zz}.
\item The rms radius of Ref.~\cite{Bernauer:2010wm} includes the
correct and complete errors of a detailed error analysis compatible
with the book mentioned above.
\end{itemize}

\begin{table}[hbt]
\caption{Various moments of the charge distribution for the standard
dipole form and the assembly of the fits of Bernauer {\it et al.}\ and
Arrington {\it et al.}. All units are in powers of fm. Note that
$\langle r \rangle_{(2),em}$ represents the so called Zemach radius
calculated with the corresponding Eq.~(\ref{eq:3}) for the electric and
magnetic form factors. An additional numerical uncertainty is given
for the highest moments.} \label{tab:2}
\begin{center}
\begin{tabular}{l|r@{.}l|r@{.}l}
 & \multicolumn{2}{c|}{standard} & \multicolumn{2}{c}{Bernauer-} \\
 & \multicolumn{2}{c|}{dipole}   & \multicolumn{2}{c}{Arrington} \\ \hline
$\langle 1/r^2 \rangle$      &  9&117  &  8&100(19)  \\
$\langle 1/r \rangle$        &  2&135  &  2&0666(24) \\
$\langle r \rangle$          &  0&7026 &  0&7381(17) \\
$\langle r^2 \rangle$        &  0&6581 &  0&774(8)   \\
$\langle r^3 \rangle$        &  0&7706 &  1&16(4)    \\
$\langle r^4 \rangle$        &  1&083  &  2&59(19)(04)\\
$\langle r^5 \rangle$        &  1&775  &  8&0(1.2)(1.0)\\
$\langle r^6 \rangle$        &  3&325  &  29&8(7.6)(12.6)\\
$\langle \log(r) \rangle$    & -0&5289 & -0&4944(12) \\
$\langle r^2\log(r) \rangle$ &  0&0358 &  0&169(12)  \\
$\langle r \rangle_{(2),em}$  &  1&025  &  1&045(4)
\end{tabular}
\end{center}
\end{table}

\section{Conclusion} \label{sec:5}

Having disproven the proposal of De~R\'{u}jula by the experiments again,
the question arises where else one could find a reconcilement of the
two disagreeing radii. One question is, of course, whether the QED
corrections are complete or whether some higher order contributions
are not yet sufficiently well studied. The most recent, comprehensive
discussion of these possible deficiencies can be found in the paper by
Edith Borie \cite{Borie:2004fv}. One might add that it could be
necessary that the distorted relativistic wave functions had to be
treated in a perturbative sum over multiphoton exchanges. The
convergence of such sums is not proven in QED. Such a calculation
may imply the solution of a Bethe-Salpeter equation and appears
hardly possible \cite{Eides:2000xc}. However, further theoretical
studies may finally show that the approximations made so far are
adequate.

In this case, one had to go to fancier ideas but still within the
framework of the standard model. We mention the so called ``hadronic
corrections'' as the polarisation of the proton or mesonic loops. We
could, for example, assume a further particle-antiparticle fluctuation
in the point Coulomb field, i.e.\ beyond the $e^+e^-$ and $\mu^+\mu^-$
pairs included in the calculations of the Lamb shift so far. However,
inserting the $\pi$ mass, the lightest hadron known, in the
nonrelativistic approximation of Pachucki \cite{Pachucki:1999zza} gives
a small contribution of -0.0095\,meV only, already taken into account
in Eq.~(\ref{eq:5}) \cite{Eides:2000xc}. If we assign the difference
of the two determinations of the radius of 0.038\,fm fully to the energy
difference of the $2S-2P$ point nucleus Lamb shift we get a shift of
-0.341\,meV.  The mass of an electrically charged particle-antiparticle
pair producing such a Lamb shift would be 23\,MeV. No free particle with
this mass is known.

However, quantum mechanics demands fluctuations of quarks-antiquarks
also in the Coulomb field and one might identify these particles with
quarks. It is not evident that these pairs have to form the asymptotic
Goldstone Boson, the pion, resulting from the breaking of chiral
symmetry of QCD, if this is produced dynamically, i.e.\ including
gluonic components. One could think of the vacuum in terms of a Fock
state expansion and consider the contribution to the vacuum polarisation
as a sum of terms yielding a low effective mass. Inserting the quark
charges and assuming $2\,m_{up}=m_{down}$ one gets
$m_{up,\textrm{effective}}=\unit[28]{MeV/c^2}$. Considering the small
energy scales in muonic hydrogen this does not unreasonably compare to
the current quark masses $m_{up}\approx\unit[3]{MeV/c^2}$ and
$m_{down}\approx\unit[6]{MeV/c^2}$ at the $\unit[2]{GeV}$ scale
\cite{Manohar:2008zz}. If such QCD loops are present one has to
investigate their influence on the QED corrections for the hyperfine
splitting of muonic hydrogen as well.

Ironically, one could revert the interpretation of the muonic hydrogen
experiment if we assume that the QED calculations are sufficiently
exact. By inserting the precise radius from the electronic
experiments and the safe Zemach moments presented in this paper, one
can determine the polarisation in the Coulomb field by quark loops or
other hadronic, possibly dual, corrections at a very low $Q^2$ scale.
This idea needs a thorough theoretical study going beyond the scope of
this paper.

\section*{Acknowledgement}
We thank Ulrich Jentschura for drawing our attention to the need for a
better determination of the Zemach moments and Stefan Scherer and Marc
Vanderhaeghen for discussions on the theoretical aspects. J\"{o}rg Friedrich 
and Ingo Sick have pointed to a numerical inconsistency. This work
was supported by the Collaborative Research Center SFB443 of the
Deutsche Forschungsgemeinschaft.


\end{document}